\documentclass[a4paper,12pt]{article}
\usepackage[a4paper,top=2cm,bottom=2cm,left=4cm,right=1.5cm]{geometry}
\usepackage[T1]{fontenc}
\usepackage[english]{babel}
\usepackage[utf8]{inputenc}
\usepackage{amsmath}
\usepackage{amsfonts}
\usepackage{bm}
\usepackage{xcolor}
\usepackage{graphicx}

\title{Dissipative systems with nonlocal delayed feedback control}
\newcommand{\mytitle}{Dissipative systems with nonlocal delayed feedback control}
\newcommand{\myauthors}{Josua Grawitter, Reinier van Buel, Christian Schaaf, and Holger Stark}
\usepackage[pdftex,
            pdfauthor={\myauthors},
            pdftitle={\mytitle}]{hyperref}
\usepackage{comment}
\usepackage{setspace}
\let\OLDthebibliography\thebibliography
\renewcommand\thebibliography[1]{
  \OLDthebibliography{#1}
  \setlength{\parskip}{0pt}
  \setlength{\itemsep}{0pt plus 0.3ex}
}

\newcommand{\dInt}{\,\mathrm{d}}
\newcommand{\tp}{{t^\prime}}
\newcommand{\systemdelay}{{\tau_\mathrm{s}}}
\newcommand{\controldelay}{{\tau_\mathrm{c}}}
\newcommand{\controlstrength}{\kappa}

\renewcommand{\vec}[1]{\bm{#1}}
\renewcommand{\dot}[1]{\frac{\partial {#1}}{\partial t}}

\usepackage{lmodern}

\usepackage{microtype}

\begin{document}

\noindent
June~12, 2018

\vspace{0.6cm}\noindent
{\LARGE \textbf{\mytitle}}

\vspace{0.6cm}\noindent
{\textit{\myauthors}}

\vspace{0.6cm}
\noindent
Institut für Theoretische Physik,
Technische Universität Berlin,
10623 Berlin, Germany
%\vspace{0.1cm}

\noindent\rule{\textwidth}{0.4pt}

%NJP: must be understandable for broad audience; demonstrate impact

\subsection*{Abstract}
% [problem]
We present a linear model, which mimics the response of a spatially extended dissipative medium to a distant perturbation, and investigate its dynamics under delayed feedback control.
% [approach]
The time a perturbation needs to  propagate to a measurement point is captured by an inherent delay time (or latency).
% [results]
A detailed linear stability analysis demonstrates that a non-zero system delay acts destabilizing on the otherwise stable fixed point for sufficiently large feedback strengths.
The imaginary part of the dominant eigenvalue is bounded by twice the feedback strength.
In the relevant parameter space it changes discontinuously along specific lines when switching between branches of eigenvalues.
When the feedback control force is bounded by a sigmoid function, a supercritical Hopf bifurcation occurs at the stability-instability transition.
The frequency and amplitude of the resulting limit cycles respond to parameter changes like the dominant eigenvalue.
In particular, they show similar discontinuities along specific lines.
These results are largely independent of the exact shape of the sigmoid function.
Our findings match well with previously reported  results on a feedback-induced instability of vortex diffusion in a rotationally driven Newtonian fluid [Zeitz~M, Gurevich~P, and Stark~H 2015 \textit{Eur.~Phys.~J.~E}~\textbf{38} 22].
% impact
Thus, our model captures the essential features of nonlocal delayed feedback control in spatially extended dissipative systems.

\noindent\rule{\textwidth}{0.4pt}

\section{Introduction}
\label{sec:intro}

%NJP: must be understandable for broad audience; demonstrate impact

\noindent
Many phenomena in soft matter science require exciting a dissipative material.
From mixing two liquids~\cite{whitesides_origins_2006}, sorting colloids~\cite{zhang_fundamentals_2016,schaaf_inertial_2017}, controlling reaction rates~\cite{samiei_review_2016} and heat transport in microfluidic devices~\cite{li_creation_2009},  to fluid optics~\cite{whitesides_origins_2006} and  spiral patterns in liquid crystals~\cite{tran_change_2017}, soft matter systems often display their most useful or interesting properties under external stresses.
Several control and driving schemes have already been applied to these systems, including optimal control~\cite{prohm_optimal_2013}, hysteresis control~\cite{zeitz_feedback_2015}, and time-delayed feedback~\cite{luethje_control_2001,zeitz_feedback_2015,lospichl_time_2018}.
These methods sense the characteristic response of a material and adapt their control to it.
Here, we focus on time-delayed feedback control of a linear model system with internal delay, which captures the essential features of how soft matter systems respond to such a feedback scheme.

Time-delayed feedback is a control strategy that was originally proposed by Pyragas to control chaotic systems and stabilize unstable periodic orbits~\cite{pyragas_continuous_1992,schoell_handbook_2008}.
It has since been applied to various dynamical systems, such as lasers~\cite{otto_modeling_2010,pisarchik_control_2014,munelly_onchip_2017} and neural networks~\cite{schoell_time_2009}.
The method falls into the broader category of closed-loop control schemes because its control force depends purely on the present and past states of the controlled system.
Often, time-delayed feedback is called a \emph{noninvasive} control scheme, as its stabilizing force ideally vanishes in a stabilized state~\cite{schoell_time_2010}.

Earlier theoretical studies~\cite{baba_giant_2002,zeitz_feedback_2015,lospichl_time_2018} and experiments~\cite{luethje_control_2001,kiss_tracking_2006} applied delayed feedback to the spatiotemporal dynamics of specific systems.
While several investigations of purely temporal systems with an intrinsic latency have focused on stabilizing unstable foci~\cite{just_influence_1999,hoevel_control_2005,wuensche_noninvasive_2008}, in one earlier study on vortex diffusion at low Reynolds numbers an oscillating fluid flow in a circular geometry was initiated by delayed feedback~\cite{zeitz_feedback_2015}.
Thus, time-delayed feedback used \emph{invasively} can also destabilize stable fixed points and thereby potentially create novel nonequilibrium states in soft matter systems.

\begin{figure}[tp]
\centering
\includegraphics[width=8cm]{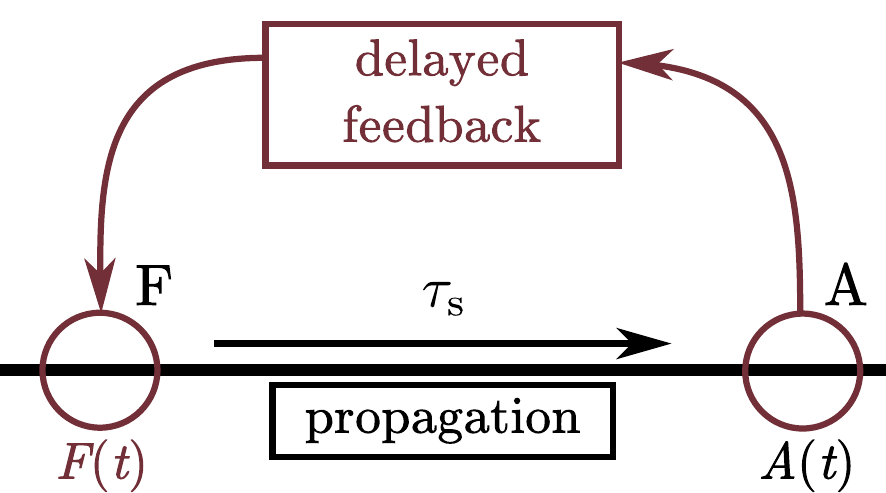}
\caption{%
Schematic of the model with nonlocal delayed feedback.
The process~$A(t)$ measured at location~A feeds back to the time-delayed control force~$F(t)$ at location, the response of which then propagates during time~$\systemdelay$ to location~A.
}
\label{sketch}
\end{figure}

In spatially extended systems the response to a control force at location~F needs the system or intrinsic delay time to propagate to a distant location~A.
There, it is sensed  and then fed back into the system at the first location~F (see the schematic in Fig.~\ref{sketch}).
To mimic this situation, we propose a linear dissipative model system, with the inherent system delay time appearing in the response function.
We then apply additional time-delayed feedback  control and concentrate on the case, where the system remains stable for vanishing system delay.
We perform a detailed linear stability analysis and demonstrate how a nonzero system delay acts destabilizing on the otherwise
stable fixed point.
Typically, the response of a physical system to an external stimulus is bounded by nonlinearities.
To mimic such a behavior, we introduce a sigmoid function to limit the strength of the feedback control force.
As a result, stable limit cycles appear in the unstable parameter regions.
While their amplitudes depend on the specific realization of the sigmoid as hyperbolic tangent, algebraic sigmoid, and ramp function, their frequencies are similar.
Both, the stability-instability transition and the appearance of limit cycles, match to the findings in Ref.~\cite{zeitz_feedback_2015}.
This suggests that our model captures the essential features of a spatially extended dissipative systems  when subjected to nonlocal delayed feedback.

We present the linear dissipative model in Sec.~\ref{sec:fixedpoint}, derive the characteristic function of its fixed point, and describe our numerical methods.
In Sec.~\ref{sec:linear_stability} we investigate and discuss the linear stability of the fixed point.
In Sec.~\ref{sec:limitcycles} we modify the feedback term by the sigmoid function such that its absolute value is limited and study the stable limit cycles arising from this modification.
Finally, we summarize our findings and conclude in Sec.~\ref{sec:conclusions}.

\section{Linear model with nonlocal delayed feedback}
\label{sec:fixedpoint}

\subsection{Derivation}
\label{sec:model}

We derive a simple model for a dissipative physical process $A(t)$, which is driven by an external, nonlocal force~$F(t)$.
Generally, the linear response of $A(t)$ to $F(t)$ is written as
\begin{equation}
A(t) = \int\limits_{-\infty}^{\infty} \chi(t - \tp) F(\tp) \dInt\tp ~,
\label{eq_response}
\end{equation}
where the response function~$\chi(t)$ characterizes the physical system completely.
If $F$ acts at some distance from the position where $A$ is observed, there will be a system delay (or latency) quantified by time $\systemdelay$, before $A$ is affected by $F$.
We describe this causal link in $\chi$ using the Heaviside function $\Theta(t-\systemdelay)$.
Furthermore, in dissipative systems the response to some perturbation often decays exponentially with rate $\alpha$ and the response function becomes
\begin{equation}
\chi(t) = \Theta(t - \systemdelay) \exp[-\alpha (t - \systemdelay)] \; .
\label{eq_response_approx}
\end{equation}
Substituting $\chi(t)$ into Eq.~(\ref{eq_response}) and differentiating both sides with respect to $t$, we find
\begin{equation}
	\dot{A(t)} = -\alpha A(t) + F(t - \systemdelay) \, .
    \label{eq_diff_start}
\end{equation}
Following Pyragas~\cite{pyragas_continuous_1992}, we now implement for $F(t)$ delayed feedback control with delay time $\controldelay$, control strength $\controlstrength$, and a constant external force $F_0$,
\begin{equation}
	F(t) = F_0 - \controlstrength [A(t) - A(t - \controldelay)] \, ,
\end{equation}
and substitute the expression into Eq.~(\ref{eq_diff_start}):
\begin{equation}
\dot{ A(t)} = -\alpha A(t) + F_0
- \controlstrength \left[ A(t - \systemdelay) - A(t - \systemdelay - \controldelay) \right] \, .
\end{equation}

This equation has one fixed point~$A^*$, where $\partial A / \partial t= 0$, which is determined by the first two terms because the delayed control term vanishes for constant solutions,
\begin{equation}
	A^* = \frac{F_0}{\alpha} \, .
\end{equation}
We nondimensionalize time~$t$ and $\controlstrength$ with $\alpha$, force~$F_0$ with $\alpha A^*$, and $A$ with $A^*$ to obtain
\begin{equation}
	\dot{A(t)} = - A(t) + 1 - \controlstrength \left[ A(t - \systemdelay) - A(t - \systemdelay - \controldelay) \right] \, .
\label{eq.DDE}
\end{equation}
In the following we study this form of the delay differential equation (DDE).

Note that for $\controlstrength<0$ the delayed feedback term acts to destabilize $A$, because when $A(t)>A(t-\controldelay)$ the feedback term gives a positive contribution to $\partial A / \partial t$.
Already for $\systemdelay=0$, this destabilizes the system for sufficiently large $|\controlstrength|$.
In the following we take $\controlstrength\geq0$ to explore the role of $\systemdelay$ in the model.

\subsection{Characteristic function}
Within control theory Eq.~(\ref{eq.DDE}) is categorized as a closed loop system with delayed feedback.
Because such systems can become unstable~\cite{niculescu_delay_2010}, we investigate the linear stability of the fixed point at $A = 1$ by introducing the perturbation ansatz
\begin{equation}
	A(t) = 1 + \varepsilon \exp(\lambda t)
\end{equation}
with $|\varepsilon| \ll 1$ and complex eigenvalues $\lambda\in \mathbb{C}$ into Eq.~(\ref{eq.DDE}), which gives a transcendental equation for $\lambda$,
\begin{equation}
	\lambda = - 1 - \controlstrength \mathrm{e}^{-\lambda\systemdelay} \left(1 - \mathrm{e}^{-\lambda\controldelay}\right) \, .
\end{equation}
In the following we solve this equation numerically and study the properties of its solutions.

The eigenvalues~$\lambda$ are roots of the characteristic function $g(\lambda)$,
\begin{equation}
g(\lambda) = \lambda + 1
+ \controlstrength\, \mathrm{e}^{-\lambda \systemdelay} \left(
	1 - \mathrm{e}^{-\lambda \controldelay}
\right) \, .
\label{eq_characteristic}
\end{equation}
Note that they cannot be expressed using the Lambert $W$ function, as is usually possible for time-delay differential equations~\cite{schoell_time_2010}, due to the double delay terms.
However, we can infer some properties of the roots from the structure of $g$.
First, with each complex $\lambda$ also its complex conjugate $\bar \lambda$ is a root of $g(\lambda)$ since $\controlstrength$, $\systemdelay$, and $\controldelay$ are real parameters.
Second, in appendix~\ref{sec:box} we show that any root $\lambda$ of $g$ with $\mathrm{Re}(\lambda) > 0$ has a nonzero imaginary part bounded by $2\controlstrength$, $|\mathrm{Im}(\lambda)| < 2\controlstrength$.
Thus any eigenvalue associated with an unstable fixed point has a nonzero imaginary part, which is bounded by control strength.
Since the imaginary part is the oscillation frequency, we conclude that fast oscillations require sufficiently large control strengths.

\subsection{Numerical methods}
\label{sec:numerics}

\subsubsection{Root finding}
In general, the roots of the characteristic function $g(\lambda)$ cannot be determined analytically.
Therefore, we use a numerical root finding algorithm to find the dominant eigenvalue in our stability analysis, i.e., the eigenvalue with the largest real part for one set of system parameters.
As a prerequisite we need to restrict the complex plane to a finite box, which is guaranteed to contain the dominant eigenvalue.
In appendix~\ref{sec:box} we prove the dominant eigenvalue for any parameter combination is contained in the compact complex region $\mathcal{B}\subset \mathbb{C}$.
\begin{align}
\mathcal{B}  =  \left\{ \vphantom{\frac{1}{\systemdelay}}\right.
\lambda \in \mathbb{C} |
-1 &\leq \mathrm{Re}(\lambda) \leq \max\left[0,\frac{1}{\systemdelay}
W\left( 2\controlstrength \systemdelay \mathrm{e}^\systemdelay \right) - 1\right]~,
\nonumber \\
0 &\leq \left.\vphantom{\frac{1}{\systemdelay}}\mathrm{Im}(\lambda) \leq \controlstrength \exp [-\systemdelay \mathrm{Re}(\lambda)] \left\{ 1 + \exp[-\controldelay \mathrm{Re}(\lambda)] \right\}
\right\}
\end{align}
We search a rectangular superset of $\mathcal{B}$ (see appendix) using an interval Newton method, which provably finds all complex roots of $g(\lambda)$ in $\mathcal{B}$.
For example, it is described in Ref.~\cite{moore_introduction_2009} and implemented in Ref.~\cite{interval_2018}.

\subsubsection{Trajectories}
We also calculate the time evolution of $A(t)$ to investigate its long-time behavior, such as limit cycles in the case of a bounded
control force.
To do so, we use the \emph{method of steps}
based on a $5$th-order Runge-Kutta method~\cite{tsitouras_runge_2011} and the $4$th-order Rosenbrock method \textit{RODAS} \cite{hairer_solving_1991} when using the ramp function to bound the control force (see Sec.~\ref{sec:limitcycles}).
Both methods are implemented in the software package \textit{DifferentialEquations.jl} \cite{rackauckas_differential_2017,ordinary_2018,delay_2018}.

The method of steps treats delay terms by splitting the domain of integration into a sequence of time intervals, so that on each interval the delayed values of the dynamic quantities are fully known.
In our case [see Eq.~(\ref{eq_dynamics_bounded}) in Sec.~\ref{sec:limitcycles}] the method initially requires a history function describing $A(t)$ for all times $-(\systemdelay + \controldelay) < t \leq 0$.
The history function is used to integrate over the interval $0 < t \leq \systemdelay$ because $A(t-\systemdelay)$ [and $A(t-\systemdelay-\controldelay)$] are then known until $t=\systemdelay$.
For remaining intervals $n\systemdelay < t \leq (n+1)\systemdelay$ (with $n\in\mathbb{N}$) integration continues using $A(t)$ from the previous intervals.

All numerical calculations are performed using the \emph{Julia} programming language~\cite{bezanson_julia_2012,bezanson_julia_2017} and all plots are created using the \emph{matplotlib} package~\cite{droettboom_matplotlib_2018}.

\section{Linear stability analysis}
\label{sec:linear_stability}
\subsection{Stability for vanishing delays}
\label{subsec.vanish}

In the limiting case of vanishing system delay, $\systemdelay\to0$, the fixed point is always stable.
To prove this, we set $\systemdelay=0$ and take the real part of $g$:
\begin{equation}
\mathrm{Re}[g(\lambda)] =
\mathrm{Re}(\lambda) + 1 + \controlstrength
- \controlstrength  \mathrm{e}^{-\controldelay\mathrm{Re}(\lambda)} \cos\left[\controldelay\mathrm{Im}(\lambda)\right] \, .
\label{eq.Re_g}
\end{equation}
We prove by contradiction:
Suppose that $\mathrm{Re}(\lambda) \ge 0$ together with $\controlstrength,\controldelay > 0$.
Then the final term in Eq.~(\ref{eq.Re_g}) is the only (potentially) negative contribution necessary to obtain $\mathrm{Re}(g) = 0$.
However, because the absolute value of this term is always smaller than or equal to $\controlstrength$, roots with $\mathrm{Re}(\lambda) \ge 0$ cannot exist.
Therefore, all eigenvalues have $\mathrm{Re}(\lambda) < 0$ and the fixed point must be stable for $\systemdelay \to 0$.

In the limit of vanishing control delay, $\controldelay \to 0$, the fixed point is stable because the characteristic function
\begin{equation}
\lim_{\controldelay \to 0} g(\lambda)
= \lambda + 1
\end{equation}
only has one (real) root at $\lambda = -1$.
This is also clear from Eq.~(\ref{eq.DDE}) since the delay term vanishes completely.

\subsection{Stability-to-instability transition}

\begin{figure}[t]
\centering
\includegraphics[width=11cm]{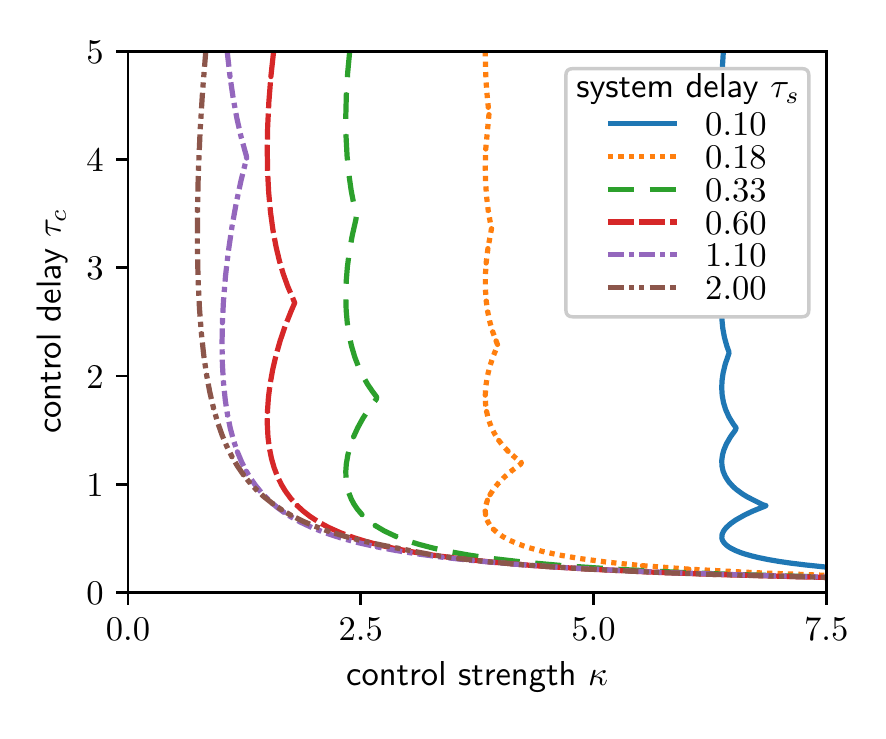}
\caption{%
Stability-instability transition curves in the $\controlstrength$--$\controldelay$ plane for various $\systemdelay$ with the stable region to the left of each curve.
}
\label{plot_transitions}
\end{figure}

A fixed point is unstable if $\mathrm{Re}(\lambda) > 0$ for its dominant eigenvalue~$\lambda$.
Based on numerical calculations of the dominant eigenvalues over a wide set of parameter combinations, we make several observations, which we report here.
For each $\systemdelay > 0$ and $\controldelay > 0$, there is a critical control strength $\controlstrength_\mathrm{crit}(\systemdelay, \controldelay)$ determined by $\mathrm{Re}(\lambda) = 0$ beyond which the fixed point is unstable for all $\controlstrength > \controlstrength_\mathrm{crit}$.
The $\controlstrength_\mathrm{crit}$ form a manifold in parameter space; cross sections for various $\systemdelay$ are displayed in Fig.~\ref{plot_transitions}.
For $\controldelay \to 0$ the critical value $\controlstrength_\mathrm{crit}$ diverges like  $\controldelay^{-1}$.
In this limit $g(\lambda)$ is approximated to linear order in $\controldelay$ as
\begin{equation}
g(\lambda) = \lambda + 1
+ \controlstrength \controldelay\lambda\, \mathrm{e}^{-\lambda \systemdelay} \, ,
\end{equation}
from which follows that any root~$\lambda$ stays the same as long as the product~$\controlstrength \controldelay$ remains constant.
For increasing~$\controldelay$ specific values of $\controlstrength_\mathrm{crit}$ exist, where the unstable eigenvalue switches from one branch to another.
We observe the resulting cusps become less prominent as $\controldelay$ increases and eventually $\controlstrength_\mathrm{crit}$ as a function of $\controldelay$ approaches a constant value for each $\systemdelay$.
Notably, these cusps imply that for $\controlstrength$ close to the cusp value alternating regions of stability and instability occur as $\controldelay$ increases.
We also observe that for small $\systemdelay$ the cusps are closer to each other w.r.t.~$\controldelay$ and $\controlstrength_\mathrm{crit}$ diverges with decreasing $\systemdelay$.
From these observations and the findings of Sec.~\ref{subsec.vanish}, we conclude that only the combination of \emph{both} nonzero delays causes the instability.

\begin{figure*}[t]
%\centering
\includegraphics[width=15.5cm]{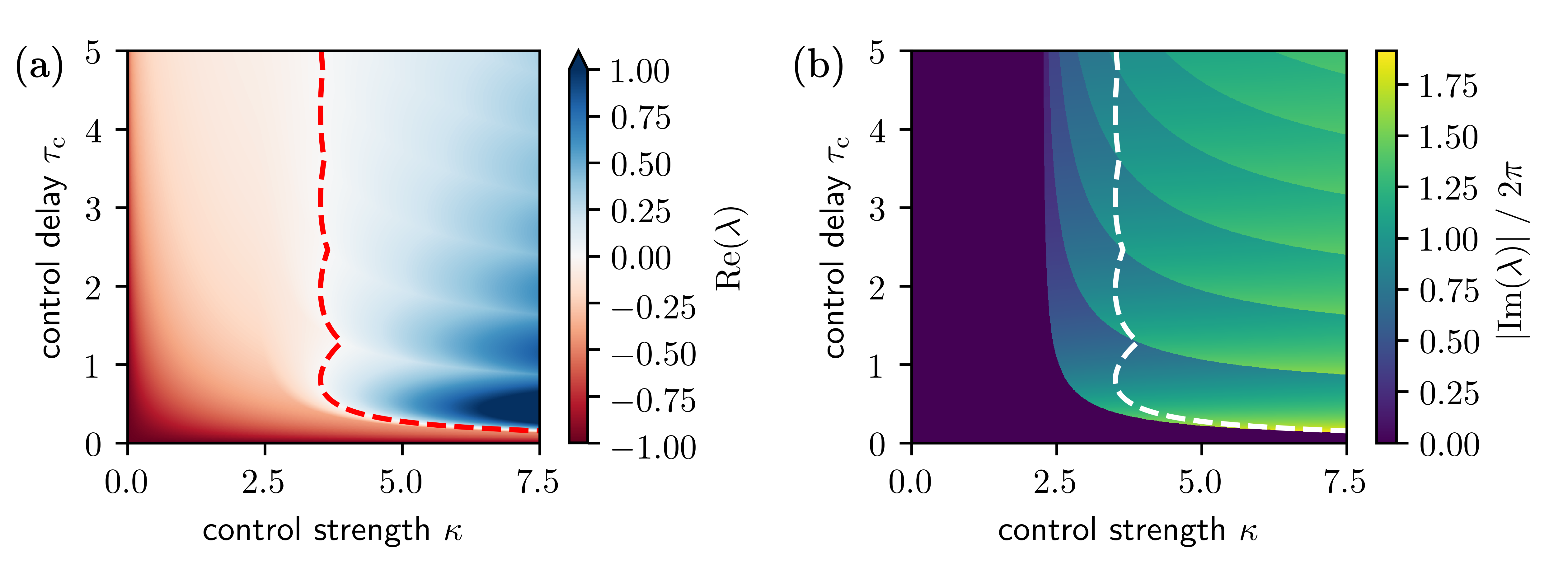}
\caption{Dominant eigenvalue $\lambda$ color-coded in $\controldelay$--$\controlstrength$ plane together with stability-instability transition curve~$\controlstrength_\mathrm{crit}$ (dashed red/white lines) for system delay $\systemdelay = 0.2$: (a) real part and (b) imaginary part.
}
\label{plot_eigenvalue}
\end{figure*}

The dominant eigenvalue displays some general features (shown in Fig.~\ref{plot_eigenvalue}):
Its real part is smallest (i.e.~close to $-1$) in regions with either $\controlstrength$ or $\controldelay$ close to zero, where the control term becomes negligible.
Conversely, this implies that delayed feedback slows down the decay of individual modes.
The imaginary part of the dominant eigenvalue is nonzero and displays clear discontinuities w.r.t.~$\controldelay$ and $\controlstrength$ in the unstable region and close to the transition curve $\controlstrength_\mathrm{crit}$ [see Fig.~\ref{plot_eigenvalue}(b)].
Since also $\mathrm{Re}(\lambda) = 0$ at the transition, there is a supercritical Hopf bifurcation.
Furthermore, as $\controldelay$ increases the imaginary part repeatedly decreases and then jumps to a larger value when the dominant eigenvalue switches to another branch.
Figure~\ref{plot_eigenvalue}(a) shows that the jump lines correspond to valleys in the real part of the dominant eigenvalue.

We furthermore note the similarities between these dominant eigenvalues and eigenvalues described in Ref.~\cite{zeitz_feedback_2015} for the specific case of delayed feedback control applied to vortex diffusion in a circular geometry.
The characteristic function for that system is derived by solving the spatiotemporal problem explicitly.
It contains Bessel functions, which play a similar role as the exponentials in Eq.~(\ref{eq_characteristic}).
Furthermore, the diffusive response function [compare Eq.~(\ref{eq_green_diffusion}) in appendix \ref{app:mapping} with $\alpha\to0$], has an initial increase and then decays to zero, just as in our model.
We consider these similarities a strong indication that some spatiotemporal systems map well on our simplified model response functions approximately given by Eq.~(\ref{eq_response_approx}).

\section{Bounded control force}
\label{sec:limitcycles}

Typically, the response of a physical system to an external stimulus is bounded by nonlinearities.
We mimic this behavior here by implementing a bounded control force.
Bounded delayed feedback was previously studied, e.g.~to suppress \emph{overshooting} due to overcompensating control forces~\cite{benner_observing_2008},
or for hydrodynamic vortex diffusion in a circular geometry to stabilize unstable modes \cite{zeitz_feedback_2015}.

\subsection{Model and setup}

We start by modifying the DDE of Eq.~(\ref{eq.DDE}) such that the time-delayed control term is limited by a monotonically increasing odd function $\sigma(x)$ with $\sigma(\pm \infty) = \pm 1$ and $\sigma'(0) = 1$:
\begin{equation}
\dot{A(t)}
= -A(t)
  -\sigma\{
      \controlstrength [A(t - \systemdelay) - A(t - \systemdelay - \controldelay)]
\} \, .
\label{eq_dynamics_bounded}
\end{equation}
Compared to Eq.~(\ref{eq.DDE}) we set the constant force to zero,  which shifts the fixed point of the modified DDE with the bounded feedback to $A^{*} = 0$.
Linearizing around this fixed point, one recovers Eq.~(\ref{eq.DDE}) without the term $+1$.

\begin{figure}[t]
\centering
\includegraphics[width=11cm]{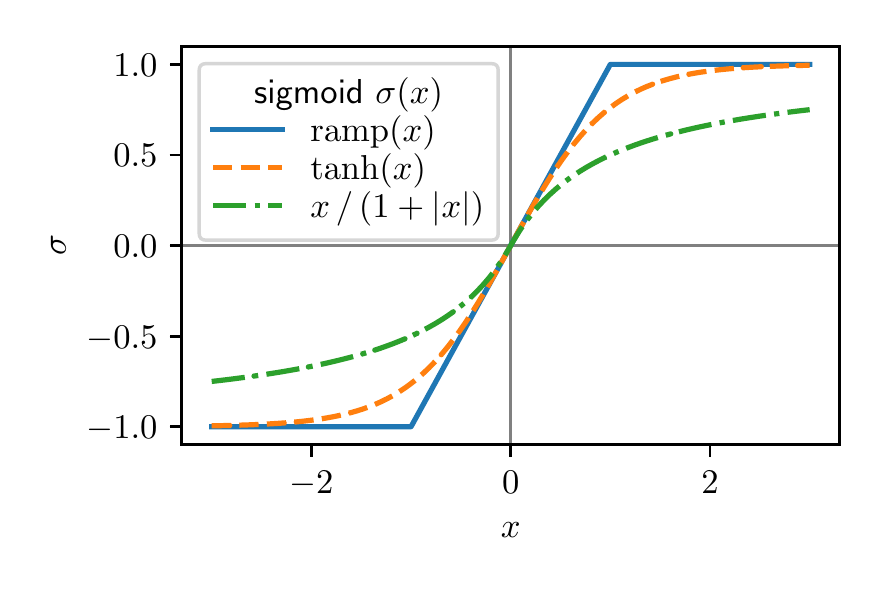}
\caption{Three realizations of a sigmoid function for implementing delayed control forces with upper and lower bounds plotted versus $x$.
Three asymptotic scalings for large $|x|$ are realized: flat (ramp), exponential ($\tanh$), and power law.}
\label{plot_sigmoids}
\end{figure}

To realize upper and lower bounds for the control forces, we consider three sigmoid functions: the hyperbolic tangent $\sigma(x) = \tanh(x)$, which approaches $\pm 1$ exponentially, an algebraic sigmoid $\sigma(x) = x (1 + |x|)^{-1}$, which approaches $\pm 1$ like a power law, and the nonsmooth ramp function $\sigma(x) = \mathrm{ramp}(x) = \min[1, \max(-1, x)]$.
All three are displayed in Fig.~\ref{plot_sigmoids}.

We solve Eq.~(\ref{eq_dynamics_bounded}) numerically as described in Sec.~\ref{sec:numerics} with the history function $A(t<0) = 0$ and a small initial perturbation, $A(t=0) = 10^{-3}$.\footnote{%
It might be possible to find analytic approximations for limit cycles (observed in the following) and their Floquet exponents using the Poincar\'e-Lindstedt method~\cite{simmendinger_analytical_1998}.
}
If the fixed point is unstable, the perturbation will grow over time, otherwise it will decay to zero.
Using this setup, we study the long-time behavior of the bounded system and its relationship to the unstable fixed point studied in Sec.~\ref{sec:fixedpoint} by calculating the time evolution of $A(t)$ until time~$T = 10^3$ and by examining the frequencies and amplitudes of the occurring limit cycles for times $0.8\,T<t<T$.

\subsection{Long-time dynamics}

\begin{figure*}[t!]
%\centering
\includegraphics[width=\textwidth]{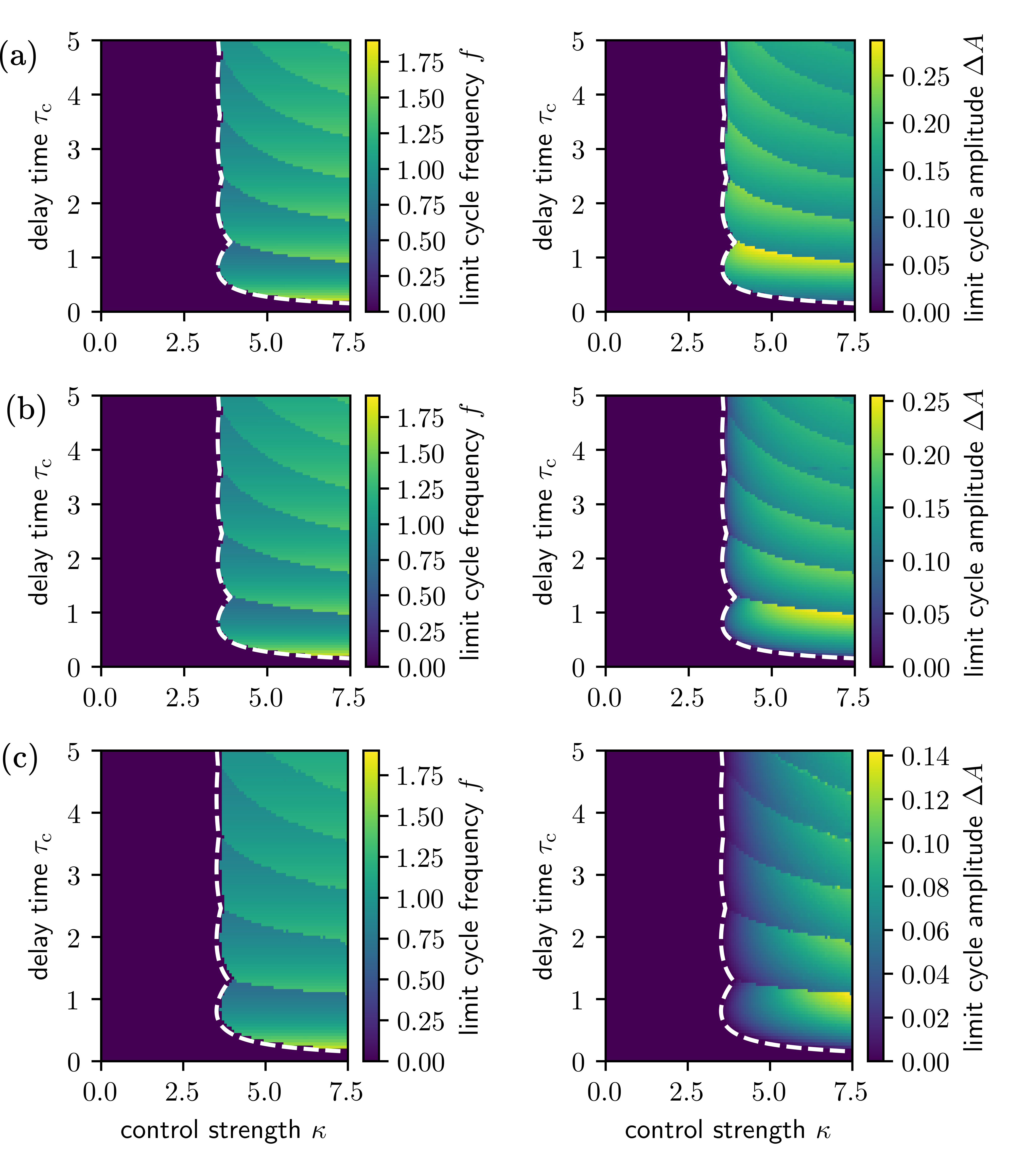}
\caption{Frequencies and amplitudes of the limit cycles resulting from the DDE of Eq.~(\ref{eq_dynamics_bounded}) color-coded in the $\controlstrength$-$\controldelay$ parameter space at $\systemdelay = 0.2$ for different realizations $\sigma$ of the bounded control force: (a) ramp sigmoid, (b) hyperbolic tangent, and (c) algebraic sigmoid.
Zero frequency (dark blue) indicates the region of stable fixed point (no oscillations).
The dashed white line indicates the transition from stable to unstable fixed point.}
\label{plot_limit_cycles}
\end{figure*}

\begin{figure}[tp]
\centering
\includegraphics[width=10cm]{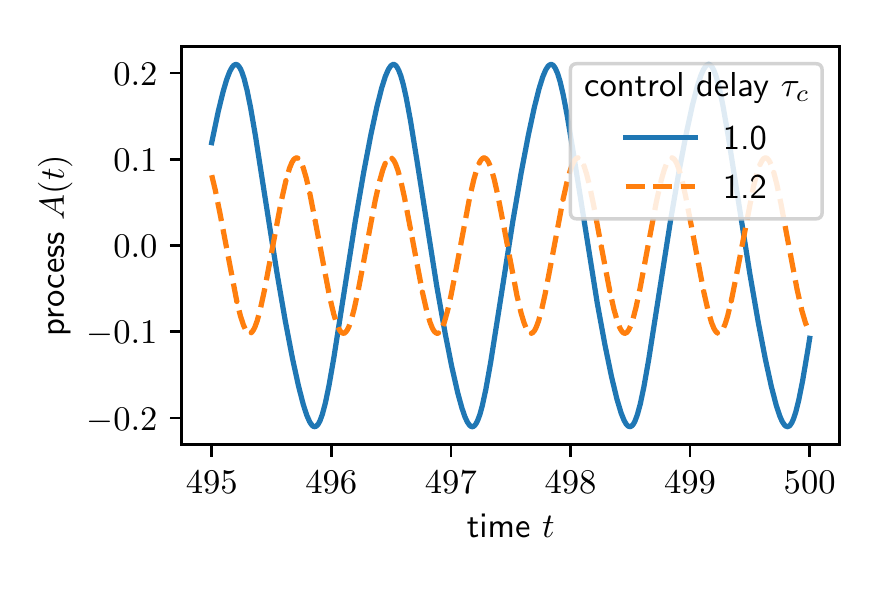}
\caption{Limit cycles of a system with feedback control bounded by the hyperbolic tangent as sigmoid function at two different control delays $\controldelay$ for $\controlstrength = 5$ and $\systemdelay = 0.2$.
The system with larger $\controldelay$ has larger periodicity but smaller amplitude.}
\label{plot_trajectories}
\end{figure}

For all three sigmoid functions $\sigma$, the bounded system displays a Hopf bifurcation, where the fixed point $A(t)=0$ becomes unstable and a limit cycle emerges (see Fig.~\ref{plot_limit_cycles}).
It is stabilized by the upper and lower bound of the control force, which would otherwise grow to infinity.
The limit cycle does not correspond to an unstable periodic orbit of the uncontrolled system as envisaged in Pyragas' original idea~\cite{pyragas_continuous_1992}, because the feedback term does not vanish when the limit cycle is reached.
Generally, the limit cycles for all sigmoid functions should converge for $\controlstrength \to \infty$ because in this limit they all approach the discontinuous step function.
In the studied parameter region, the frequencies of the observed limit cycles are determined by the imaginary part of the dominant eigenvalue at the unstable fixed point.
This becomes evident when comparing the frequency plots in the left column of Fig.~\ref{plot_limit_cycles} with Fig.~\ref{plot_eigenvalue}.
In particular, the jumps from one branch to the other agree well for both frequencies.
However, the amplitude~$\varDelta A$ of the limit cycle $A_\mathrm{lc}(t)$ defined as $\varDelta A = [\max_t\{A_\mathrm{lc}(t)\}-\min_t\{A_\mathrm{lc}(t)\}] / 2$ is generally not related to the real part of the dominant eigenvalue [compare amplitudes in the right column of Fig.~\ref{plot_limit_cycles} with Fig.~\ref{plot_eigenvalue}(a)].
It rather behaves like the frequency of the limit cycle with one difference.
It increases as $\controldelay$ grows and drops to a smaller value when the limit cycle frequency jumps to another branch.
So, at the discontinuity lines the sudden increase in frequency is accompanied by a sudden decrease in amplitude.
This is explicitly demonstrated in Fig.~\ref{plot_trajectories}.

There are some notable differences between the limit cycles generated by the three sigmoid functions bounding the control term.
Most obviously, their amplitudes (right column of Fig.~\ref{plot_limit_cycles}) behave differently at the bifurcation line: while they increase smoothly from zero for the hyperbolic tangent and algebraic sigmoid functions as in a supercritical Hopf bifurcation [rows (b) and (c)], they jump to a nonzero value for the ramp sigmoid function [row (a)].
The ramp function is a special case because it is linear up to the bounding values.
This causes the control amplitude to always reach its maximum value in the unstable region, once the transition line is crossed.
The step-like behavior could, for example, be used in experiments to accurately locate the transition line.
Furthermore, for the ramp function the amplitudes of the limit cycles are largest close to the bifurcation line, while they increase for the smooth functions when moving away from the bifurcation line with increasing $\controlstrength$.
Finally, a closer inspection shows that the discontinuity lines of the limit-cycle frequencies for the ramp function accurately track the corresponding lines in the imaginary part of the dominant eigenvalue in Fig.~\ref{plot_eigenvalue}(b).
However, there are slight deviations for the smooth sigmoid functions.

\subsection{Transient pulse trains}
\begin{figure}[tp]
\centering
\includegraphics[width=11cm]{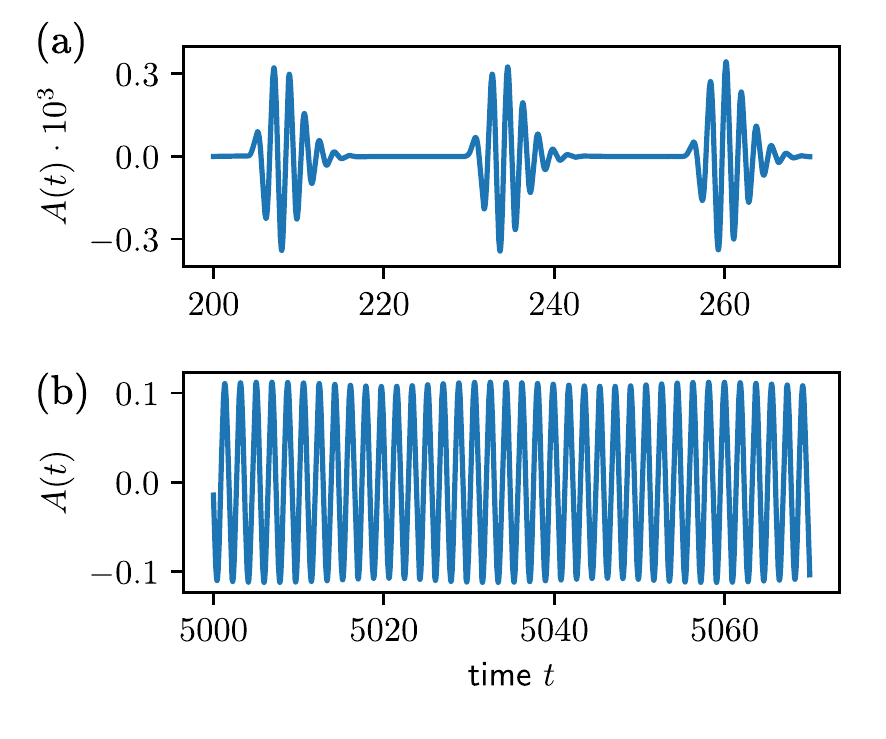}
\caption{Dynamic response of the system to feedback control bounded by the hyperbolic tangent function for $\systemdelay = 0.4$, $\controldelay = 25$, and $\controlstrength = 2.1$: (a) transient pulse trains and (b) limit cycle oscillations (note the different scales).}
\label{plot_pulses}
\end{figure}

In particular, for large control times~$\controldelay$ we observe transient pulse trains at the beginning of the dynamic evolution of our system.
They occur for stable and unstable fixed points with decaying or growing amplitude, respectively.
One example, when the fixed point is unstable, is displayed in Fig.~\ref{plot_pulses}(a).
These pulse trains grow into regular limit cycles, as shown in Fig.~\ref{plot_pulses}(b), while for stable fixed points their amplitude decays to zero.
Generally, we observe that pulse trains repeat with a periodicity given by~$\controldelay$ and their oscillation frequency is close to the imaginary part of the dominant eigenvalue~$\mathrm{Im}(\lambda)/2\pi$.

\section{Conclusions}
\label{sec:conclusions}
%\todo[inline]{NJP: must be understandable for broad audience; demonstrate impact}

Motivated by the growing interest in applying delayed feedback to soft matter systems, we have studied a linear dissipative model, which mimics nonlocal delayed feedback coupling.
To do so, we introduced an inherent system delay in the response function, which represents the time a perturbation needs to propagate to a distant location.
Our results provide an indication how nonlocal delayed feedback in a spatially extended system determines its dynamics and helps to classify the observed spatiotemporal response.

In our investigations we concentrated on the case where the system remains in its stable fixed point when the intrinsic delay is not present.
Turning on the intrinsic delay, the feedback-driven system becomes unstable for sufficiently large control strengths.
We are able to show that the absolute imaginary part of the dominant eigenvalue is bounded from above by  twice the feedback strength.
Stability-instability transition curves and the imaginary part of the dominant eigenvalue match well  with the findings reported in Ref.~\cite{zeitz_feedback_2015}, where delayed feedback was applied to vorticity diffusion of a Newtonian fluid in a circular geometry at low Reynolds numbers.
This demonstrates that our simple model system captures the essential properties of the spatiotemporal dynamics in a specific system.

To further study the feedback-induced instability, we examined the long-time dynamics of our model with bounded feedback, which in linearized form yields the original system.
Realizing the bounded feedback by smooth sigmoid functions, the stability-instability transition occurs via a supercritical Hopf bifurcation.
Thus, the fixed point becomes unstable and a stable limit cycle evolves continuously.
In contrast to Pyragas' original idea~\cite{pyragas_continuous_1992}, this limit cycle does not correspond to an unstable periodic orbit of the uncontrolled system, but is stabilized by the control force.
For all three sigmoid functions and across many parameter combinations, the frequencies of the limit cycle match well the imaginary part of the fixed point's dominant eigenvalue.
Discontinuity lines are visible, which occur when the dominant eigenvalue switches from one branch to another.
When crossing these lines by a small increase in control delay, the limit-cycle frequency jumps up and the amplitude drops sharply.
The results for the nonsmooth \emph{ramp} function  differ from the other sigmoid functions, because at the Hopf bifurcation the amplitude of the limit cycle jumps to a nonzero value.
In addition, the discontinuity lines of the limit-cycle frequencies accurately track the corresponding lines for the imaginary part of the dominant eigenvalue of the linearized system.
Both features predestine the \emph{ramp} sigmoid function to accurately determine the stability-instability transition in an experimental system.

The linear dissipative model presented in this article with its intrinsic delay time helps to classify and understand the spatiotemporal response of a spatially extended system subject to nonlocal delayed feedback.
Our work demonstrates that this model already exhibits complex and non-trivial behavior.
It also provides an example for double delay systems, which have recently attracted attention~\cite{li_impulsive_2013,wang_double_2017,brunner_spatio_2017}.
In future studies we will apply delayed feedback to specific nonlinear soft matter systems such as photoresponsive fluid interfaces~\cite{grawitter_feedback_2018} and viscoelastic flow in Taylor-Couette geometry~\cite{buel_something_2018}.
The work presented in this article will help us to categorize the observed spatiotemporal dynamics.

\subsection*{Acknowledgements}

We acknowledge financial support from DFG (German Research Foundation) \textit{via} International Research Training Group~1524 and Collaborative Research Center~910.

\appendix

\section{A physical system with intrinsic delay}
\label{app:mapping}
We consider a diffusion-reaction equation of a substance with density $\rho(\vec r, t)$, which decays at a rate $\alpha$ in two dimensions:
\begin{equation}
\partial_t\rho(\vec r,t) = D \nabla^2\rho(\vec r,t)- \alpha\rho(x,t) \, ,
\end{equation}
where $D$ is the diffusion constant.
As initial condition at $t=0$ we choose a delta peak located at $\vec r=0$, ${\rho(\vec r,0)=N\delta(\vec r)}$.
The solution of this problem is given by
\begin{equation}
\rho(\vec r,t)=\frac{N}{4\pi D t}\exp\left(-\alpha t - \frac{r^2}{4D t}\right) \, ,
\label{eq_green_diffusion}
\end{equation}
where $r=|\vec r|$ is the distance from the perturbation.
The density at any point $\vec r_0\neq 0$ increases up to the time
\begin{equation}
t_0=\frac{1}{2\alpha}
\left(\sqrt{1+\frac{\alpha r_0^2}{D}}-1\right)
\end{equation}
and then decreases to zero.
Note that for pure diffusion ($\alpha\to0$) the maximum is reached at~$t_0=r_0^2(4D)^{-1}$.
Thus, a disturbance initiated at $r=0$ needs the intrinsic delay time $t_0$ to reach $\vec r_0$.
To approximate this behavior in a response function of the form given in Eq.~(\ref{eq_response_approx}), we assume a step function where the substance $\rho$ jumps to its maximum value $\rho(\vec r_0,t_0)$ and then decays exponentially,
\begin{equation}
\chi(\vec r_0,t)=\Theta(t-t_0)\rho(\vec r_0,t_0)\exp\left[-\alpha (t- t_0)\right] \, .
\end{equation}
Thus, $\systemdelay=t_0$ is the intrinsic delay time and the physical decay rate $\alpha$ of our substance enters directly the response function.

\section{Search region for dominant eigenvalues}
\label{sec:box}

We find the roots of the characteristic function $g$ numerically.
Because a numerical search on the infinite complex plane is unfeasible, we restrict our search to a bounded region which is guaranteed to contain the dominant eigenvalue, i.e., the root of $g$ with the largest real part.

First, we find an upper bound to the real part of all eigenvalues using $\mathrm{Re}[g(\lambda)] = 0$:
\begin{equation}
\frac{\mathrm{Re}(\lambda) + 1}{\controlstrength}
= \exp[-(\systemdelay + \controldelay) \mathrm{Re}(\lambda)] \cos[(\systemdelay + \controldelay) \mathrm{Im}(\lambda)]
-\exp[-\systemdelay \mathrm{Re}(\lambda)]
\cos[\systemdelay \mathrm{Im}(\lambda)]
\end{equation}
The cosines may at most change the signs of the terms such that both contribute positively.
Thus,
\begin{equation}
\frac{\mathrm{Re}(\lambda) + 1}{\controlstrength}
\leq \exp[-\systemdelay \mathrm{Re}(\lambda)]
+\exp[-(\systemdelay + \controldelay) \mathrm{Re}(\lambda)] \, .
\end{equation}
Assuming $\mathrm{Re}(\lambda) \geq 0$,
\begin{equation}
\frac{\mathrm{Re}(\lambda) + 1}{\controlstrength}
\leq 2\exp[-\systemdelay \mathrm{Re}(\lambda)] \, ,
\end{equation}
and we solve for $\mathrm{Re}(\lambda)$ using Lambert's $W$ function,
\begin{equation}
\mathrm{Re}(\lambda)
\leq \frac{1}{\systemdelay}
W\left( 2\controlstrength \systemdelay \mathrm{e}^{\systemdelay} \right) - 1 \, .
\end{equation}
With our previous assumption we have
\begin{equation}
\mathrm{Re}(\lambda)
\leq \max\left[0, \frac{1}{\systemdelay}
W\left( 2\controlstrength \systemdelay \mathrm{e}^{\systemdelay} \right) - 1\right] \, .
\end{equation}
Notably the upper bound is independent of $\controldelay$.

Second, we find a lower bound for $\mathrm{Re}(\lambda)$ such that our search region contains at least one eigenvalue.
To simplify our search, we concentrate on $\lambda \in \mathbb{R}$, which is determined by
\begin{equation}
\lambda = -1 -\exp(-\systemdelay \lambda) \left[ 1 - \exp(-\controldelay \lambda) \right] \, .
\label{eq_real_g}
\end{equation}
On the interval $(-\infty, 0]$ the l.h.s.~of this equation goes continuously from $-\infty$ to $0$ and the r.h.s.~goes continuously from $+\infty$ to $-1$.
Because both sides are continuous and strictly monotonic for $\controldelay,\systemdelay>0$, they share exactly one value in the overlap, i.e.~$-1\leq\lambda\leq0$.
Therefore, a search region with $\mathrm{Re}(\lambda)\geq-1$, which also contains the real axis, will always contain at least one (real) eigenvalue.

Third, we find an upper bound for the imaginary parts of all eigenvalues using $\mathrm{Im}[g(\lambda)] = 0$.
\begin{equation}
\mathrm{Im}(\lambda)
= \controlstrength\exp[-\systemdelay \mathrm{Re}(\lambda)] \sin[\systemdelay \mathrm{Im}(\lambda)]
-\controlstrength\exp[-(\systemdelay+\controldelay) \mathrm{Re}(\lambda)] \sin[(\systemdelay+\controldelay) \mathrm{Im}(\lambda)]
\end{equation}
Here, we simply drop the sines, assuming all terms contribute positively,
\begin{equation}
\mathrm{Im}(\lambda)
\leq \controlstrength \exp[-\systemdelay \mathrm{Re}(\lambda)] \left\{ 1 + \exp[-\controldelay \mathrm{Re}(\lambda)] \right\}
\end{equation}
The r.h.s.~depends on $\mathrm{Re}(\lambda)$ and on the interval determined for $\mathrm{Re}(\lambda)$.
It is maximal for $\mathrm{Re}(\lambda) = -1$ implying $\mathrm{Im}(\lambda)
\leq \controlstrength \, \mathrm{e}^\systemdelay [ 1 + \mathrm{e}^\controldelay ]$.
Note that the symmetry $g(\bar\lambda) = \bar g(\lambda)$ implies that we only need to search the positive half of the complex plane, i.e., $\mathrm{Im}(\lambda)\geq 0$, because all complex roots appear in conjugate pairs.

\begin{figure}[t!]
\centering
\includegraphics[width=11cm]{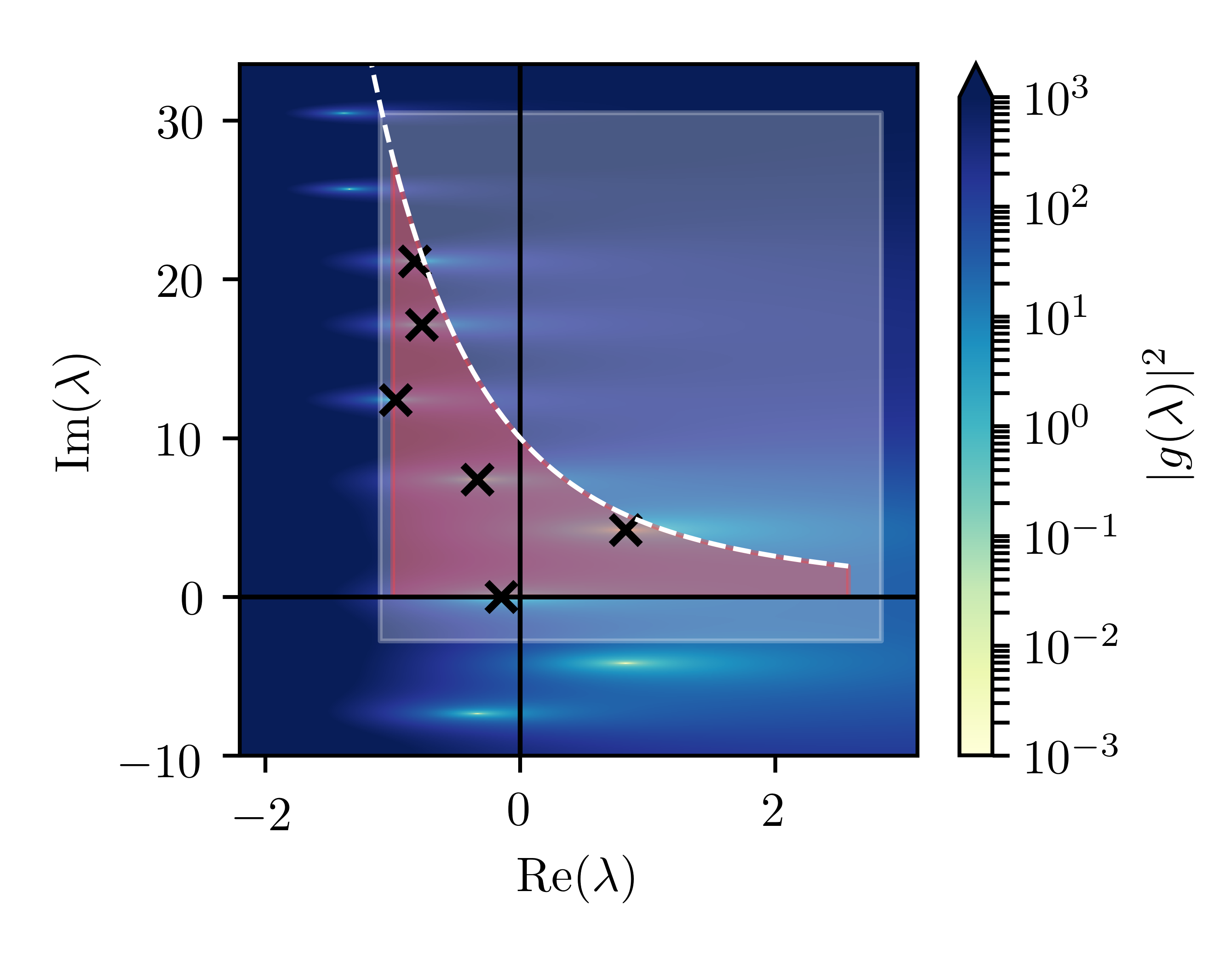}
\caption{Example of a region, which with certainty contains the dominant eigenvalue as determined in Eq.~(\ref{eq_guarantee_region}) (red, transparent), numerical search rectangle from Eq.~(\ref{eq_search_region}) (white, transparent), and numerically located eigenvalues (black crosses) for $\controlstrength=5$, $\controldelay=1$, $\systemdelay=0.4$.
The upper bound for the real parts of eigenvalues with large imaginary parts is displayed as a white, dashed line.
The background shading indicates the absolute value of the characteristic function $|g(\lambda)|$.
}
\label{plot_algorithm}
\end{figure}

In summary, we have shown that the following set of complex numbers $\mathcal{B}$ must always contain the eigenvalue with the largest real part:
\begin{align}
\mathcal{B} = \left\{\vphantom{\frac{1}{\systemdelay}}\right.
\lambda \in \mathbb{C} |
-1 &\leq \mathrm{Re}(\lambda) \leq \max\left[0, \frac{1}{\systemdelay}
W\left( 2\controlstrength \systemdelay \mathrm{e}^{\systemdelay} \right) - 1\right]~,
\nonumber \\
0 &\leq \left.\vphantom{\frac{1}{\systemdelay}}\mathrm{Im}(\lambda) \leq \controlstrength \exp[-\systemdelay \mathrm{Re}(\lambda)] \left\{ 1 + \exp[-\controldelay \mathrm{Re}(\lambda)] \right\}
\right\}
\label{eq_guarantee_region}
\end{align}
For simplicity and safety we search a bounded rectangular region which is a superset of $\mathcal{B}$:
\begin{align}
\mathcal{B} \subseteq
\left\{\vphantom{\frac{1}{\systemdelay}}\right.
\lambda \in \mathbb{C} |
-1.1 &\leq \mathrm{Re}(\lambda) \leq \max\left[0, \frac{1.1}{\systemdelay}
W\left( 2\controlstrength \systemdelay \mathrm{e}^{\systemdelay} \right) - 1.1\right]~,
\nonumber \\
-0.1 &\leq \left.\vphantom{\frac{1}{\systemdelay}}\mathrm{Im}(\lambda) \leq 1.1\, \controlstrength \mathrm{e}^{\systemdelay} \left( 1 + \mathrm{e}^{\controldelay} \right)
\right\}
\label{eq_search_region}
\end{align}
Both regions are displayed in Fig.~\ref{plot_algorithm}.

Note that the shape of $\mathcal{B}$ immediately implies that any eigenvalue $\lambda$ with $\mathrm{Re}(\lambda) > 0$ has $|\mathrm{Im}(\lambda)| < 2\controlstrength$.
Furthermore, it can be shown that $\mathrm{Re}(\lambda) > 0$ implies $|\mathrm{Im}(\lambda)| \neq 0$ because Eq.~(\ref{eq_real_g}) has no solutions for $\lambda > 0$.
In this case the l.h.s.~of Eq.~(\ref{eq_real_g}) is always positive while the r.h.s.~is always smaller than $-1$ and therefore no solution with $\mathrm{Re}(\lambda) > 0$ and $|\mathrm{Im}(\lambda)| = 0$ exists.

\newpage
%\section*{References}
\bibliographystyle{iopart-num}
\bibliography{sources}
\addcontentsline{toc}{section}{References}

\end{document}